 \documentclass[final,3p,times]{elsarticle}

\usepackage{amssymb}

\journal{Physical Letters A}

\newcommand {\be}{\begin{eqnarray}}
\newcommand {\ee}{\end{eqnarray}}

\begin{document}

\begin{frontmatter}



\title{Impossibility to describe repulsion with contact interaction}


\author{K. Morawetz$^{1,2}$, M. M{\"a}nnel$^{3}$}

\address{$^1$M\"unster University of Applied Science,
Stegerwaldstrasse 39, 48565 Steinfurt, Germany\\
$^2$International Center of Condensed Matter Physics, University of Brasilia,
70904-970, Brasilia-DF, Brazil\\
$^3$Institute of Physics, Chemnitz University of Technology, 09107 Chemnitz, Germany}

\begin{abstract}
Contact interactions always lead to attractive behavior. Arguments are presented to show why a repulsive interacting system, e.g. Bose gases, cannot be described by contact interactions and corresponding treatments are possibly obscured by the appearance of bound states. The usually used cut-offs are identified as finite range parameters.
\end{abstract}

\begin{keyword}



\PACS
03.75.Hh, 
05.30.Jp, 
05.30.-d, 
64.10.+h 

\end{keyword}

\end{frontmatter}

The contact interaction
\be
V ({\bf r})=\lambda \delta({\bf r})
\ee 
in terms of the scattering length $a_0$ \cite{DGPS99,BBHLV01} and mass $m$, i.e. $\lambda=-{4\pi\hbar^2a_0}/{m}$, is often used as a pseudopotential \cite{PS04,BW59} to describe attractive $a_0>0$ or repulsive $a_0<0$ behavior. We follow the convention for $a_0$ of \cite{GW64} which
  has an opposite sign of \cite{F94}. Especially in field-theoretical approaches it is customary to characterize the attractive/repulsive behavior of the contact interaction by the sign of the scattering length. In this note arguments are collected to show why an attractive potential is effectively generated every time when a contact interaction is used independent of the sign of the scattering length and therefore a repulsive behavior can never be reached. 
  
As a first step let us examine the off-shell T-matrix for two scattering
particles with momentum $k$ and $p$
\cite{M02} without medium effects in 3D,
\be
T_{\bf k,p,q}(\omega)=V(q)+\sum\limits_{\bf q'} V(q') {1\over \omega-{q'^ 2\over
    m}+i\eta}T_{\bf k+q',p-q',q-q'}(\omega). 
\label{tm}
\ee
The bare T-matrix from the contact interaction $\lambda \delta(x)$ reads $
T(\omega)=\lambda /( 1-\lambda J(\omega))$
with [${\rm Im} J\le 0$]
\be
J(\omega)&=&\!{m\over 2 \pi^2\hbar^3}\! \int\limits_0^{q_0}\! {q^2 d q\over m \omega\!-\!q^2\!+\!i \eta}
={-m\over 4 \pi\hbar^3} \left [ {2 q_0\over \pi}-\sqrt{-m\omega} \right
]+{\cal O}\left ({\omega\over q_0}\right ). 
\label{1}
\ee
In order to keep the scattering length finite the cut-off $q_0$ can be related to the experimental scattering length $a_0$ by the phase shift $\phi$
\be
{a_0 k }=\tan\phi ={{\rm Im} \,T({\hbar^2 k^2/m})\over {\rm Re}\, T({\hbar^2k^2/ m})}
\label{2}
\ee
which leads to an interaction strength
\be
\lambda[q_0]=-{4 \pi \hbar^2 a_0\over m} {1\over 1+{2 a_0 q_0\over \pi \hbar}}.
\label{l}
\ee  
One sees that the attractive/repulsive character of the contact interaction  follows the sign of the scattering length $a_0$ as long as the cut-off $q_0$ is below the critical value $-\pi\hbar/2 a_0$. Attractive behavior remains attractive for all cut-offs. But for repulsive scattering lengths, $a_0<0$, there appears a sign change for larger cut-offs rendering the potential effectively attractive again. This puzzling behavior is due to the regularization to reproduce the scattering length which enforces a bound state. Indeed, introducing the running coupling constant (\ref{l}) into the T-matrix leads to
\be
T(\omega)=-{4 \pi \hbar^2 a_0\over m} {1\over 1+{a_0\over \hbar} \sqrt{-m\omega}}.
\label{3}
\ee
A bound state, $\omega=-E<0$, as a pole appears in (\ref{3}) only if
$
\sqrt{m E}=-{\hbar \over a_0},
$ 
i.e. for $a_0<0$. Please note that due to the Levinson theorem \cite{GW64} the phase shift falls off from $\pi$ with increasing momenta such that $a_0<0$ and the phase shift remains positive, i.e. attractive. It is not the sign of the scattering length but the sign of the phase shift that determines the attractive or repulsive behavior. Indeed, we do not have the case of negative phase shifts, i.e. repulsion, for contact interaction. One sees this by investigating the on-shell T-matrix (\ref{3}) for $\omega=\hbar^2 k^2/m$  which reads, [see also (51a) of \cite{GW64}],
\be
T&=&-|T| {\rm e} ^{i\phi}=-{4\pi \hbar ^2 a_0\over m} {1\over 1-i {k a_0}}.
\ee
Examining imaginary and real parts lead to $\cos{\phi}=1+{\cal O}(k^2)$ and $\sin{\phi}=k a_0+{\cal O}(k^2)$.
For negative scattering lengths $a_0<0$ we obtain $\phi=\pi+{k a_0}+{\cal O}(k^3)$, i.e. a bound state and positive (attractive) phase shifts and for $a_0>0$ we obtain $\phi={k a_0}+{\cal O}(k^3)$ and no bound states but still positive phase shifts. This shows that a repulsive behavior characterized by $\phi<0$ cannot be reached. In other words by fixing the scattering length (\ref{2})
we obtain a running coupling constant (\ref{l}) just in a way that every time attraction is realized. For large cut-offs $q_0$ we have an attractive potential strength $\lambda$ in both cases $a_0\gtrless 0$.

We note that the same result (\ref{3}) appears and the above discussion applies to the regularized
potentials \cite{W90}
\be
V({\bf r})=\delta({\bf r}) \left (1+r{\partial \over \partial r} \right )
\label{reg}
\ee 
used in atomic physics, e.g. for multiple ionization problems
\cite{Ber75} or cold atoms in a trap \cite{BERW98}. Fourier-transforming into momentum space one needs to take some 
care of not interchanging integrations and obtains
\be
V({\bf q})=1-D_q=1-{\left ( q^3...\right )^{q_0}_0 \over \left ( q^3/3 \right
  )^{q_0}_0}   
\ee
where the $...$ marks the place of the term with which the potential is
multiplied. One sees that the regularization of the contact potential
(\ref{reg}) at zero distance translates into a regularization at large momentum.
We can solve the T-matrix (\ref{tm}) and obtain
\be
T(\omega)=\lambda {1-D_q\over1-\lambda J(\omega)}
\ee
with
\be
J(\omega)=\!{m\over 2 \pi^2\hbar^3}\! \int\limits_0^{q_0} dq (1-D_q) {q^2\over m \omega\!-\!q^2\!+\!i \eta}
={-m\over 4 \pi\hbar^3} \left [ -\sqrt{-m\omega} \right
]+{\cal O}\left ({\omega\over q_0}\right ). 
\label{1a}
\ee  
which leads to (\ref{3}) for the T-matrix. In other words the
regularized contact interaction performs the same regularization as the
cut-off discussed above. 

The 1D potential does not require a cut-off \cite{W90} in $J(\omega)$
and indeed leads to $\lambda=4 \pi \hbar^ 2/m a_0$ and
\be
T_{1D}(\omega)={4 \pi \hbar\over m} {\sqrt{-m \omega}\over 1+ {a_0\over
    \hbar}\sqrt{-m\omega}}.
\ee
The 2D case involves some regularization function \cite{W90} and can be
discussed analogously as above.

The fact that no repulsion is possible to describe with contact interactions irrespective of whether the scattering length is positive or negative, can be seen also by approaching the contact interaction in the limit of known potentials.
As a first example, the rectangular sphere potential with radius $R$ and scaling length $L$
\be
V(r)={\hbar^2\over 2 m L R} \Theta(R-r)\to {\hbar^2\over 2 m L } \delta(r)\quad{\rm for}\, R\to0
\ee
leads to the phase shift (problem 84 of \cite{F94})
\be
\phi=k R \left ({\tanh \sqrt{R/L}\over \sqrt{R/L}}-1\right )
\ee
which provides
$
a_0=-{1\over 3 L} R^2
$ approaching zero in the limit of contact interaction $R\to0$.
A second example is the opaque wall potential
\be
V(r)={\hbar^2\over 2 m L} \delta(r-R)
\ee
which leads to a phase shift (problem 86 of \cite{F94})
\be
\tan(x+\phi)={\tan{x}\over 1+{R\over L} {\tan{x}\over x}}
\ee
with $x=k R$ and one finds
$
a_0=-{1\over L} R^2
$ which vanishes as well in the limit of zero (hard-core like) radius $R$.
The consequence is that with a repulsive contact interaction it is not possible to generate $a_0<0$ and therefore repulsion.

There is a deep reason behind this impossibility to generate repulsive behavior with contact interactions. Eugene P. Wigner proved a theorem, that the momentum derivative of the s-wave phase shift obeys \cite{W55}
\be
{\partial \phi \over \partial k}> -R+{\sin{(2\phi+2 k R)}\over 2 k}
\label{theorem}
\ee
where $R$ is the range beyond which the potential vanishes.
Alternative derivations are presented in \cite{PC97,MEDD02}. 
Denoting 
\be
f(k)=k \cot \phi= k {{\rm Re} T \over {\rm Im} T}={1\over a_0} +\frac 1 2 r_o k^2+{\cal O} (k^3)
\label{exp}
\ee
corresponding to the effective-range expansion of the T-matrix, 
\be
{T^{-1}\left ({\hbar^2 k^2 \over m}\right )}={-m \over 4 \pi \hbar^2 } \left ( {1 \over a_0} +\frac 1 2 r_0 k^2-i k +{\cal O}(k^3) \right ),
\ee
the theorem (\ref{theorem}) translates into
\be
\!\!k {\partial f \over \partial k}\!< \!f \left [1\!-\!\cos{2 k R}\right ]
\!+\!\!\left (\!k^2\!-\!f^2\right )\! {\sin{2 k R} \over 2 k} 
\!+\!R \!\left (k^2\!\!+\!\!f^2\!\right )
\ee
and inserting (\ref{exp}) yields \cite{PC97,BCP98}
\be
r_0< 2 R \left ( 1+{R\over a_0}+\frac 1 3 {R^2\over a_0^2} \right )+{\cal O} (k).
\label{result}
\ee
Irrespective of the sign of $a_0$, the right hand side of (\ref{result}) is positive and vanishes for $R\to 0$. A vanishing potential range $R$ means exactly contact interaction and we obtain a negative effective range $r_0<0$ which shows that repulsive behavior cannot be obtained with the help of contact interactions. Only if a small cut-off $q_0< {\cal O}(1/r_0)$ is used one can preserve $r_0>0$ \cite{BCP98}. However then the effective potential is no longer a contact one but becomes an effective finite-range potential due to the introduction of the ad-hoc cut-off. This can be illustrated by the example of a separable interaction $V_{k,k'}=\lambda g_k g_{k'}$ where the form factors $g_k=\gamma^2/(\gamma^2+k^2)$ \cite{Y54} are controlled by the range parameter $\gamma$, for other form factors see \cite{MSSFL04}.
One obtains the same formula (\ref{l}) if $q_0\equiv {\pi \hbar } \gamma/4$. The cut-off $q_0$ can be therefore understood as range parameter $\gamma$ of an equivalent separable potential. Since the latter one describes a finite range, 
the cut-off introduces an effective finite-range potential. 

Summarizing, the contact interaction often used in treatments of interacting Bose gases implies, due to the necessary regularization, an attractive potential irrespective of the sign of the used scattering length
. A result is the appearance of bound states and therefore a change in the sign of the scattering length \cite{MMSL08,MMS07}
. The problems arising from the use of contact interactions have there origin in the two-particle scattering and are therefore independent of the temperature and also independent of whether the particles are bosons or fermions.

The author thanks Roland Zimmermann for clarifying discussions and Fabio L. Braghin for valuable comments. The financial support by the Brazilian Ministry of Science 
and Technology is acknowledged.

\bibliography{kmsr,kmsr1,kmsr2,kmsr3,kmsr4,kmsr5,kmsr6,kmsr7,delay2,spin,gdr,refer,sem1,sem2,sem3,micha,genn,solid,deform,bose,delay3}
\bibliographystyle{elsarticle-num}
\end{document}